\documentclass[twocolumn,showpacs,preprintnumbers,amsmath,amssymb]{revtex4}
\usepackage{amsmath}

\begin{document}

\tolerance=5000

\def\pp{{\, \mid \hskip -1.5mm =}}
\def\cL{{\cal L}}
\def\be{\begin{equation}}
\def\ee{\end{equation}}
\def\bea{\begin{eqnarray}}
\def\eea{\end{eqnarray}}
\def\beq{\begin{eqnarray}}
\def\eeq{\end{eqnarray}}
\def\tr{{\rm tr}\, }
\def\nn{\nonumber \\}
\def\e{{\rm e}}

\title{Gravity assisted dark energy dominance and cosmic acceleration}

\author{Shin'ichi Nojiri\footnote{E-mail: 
snojiri@yukawa.kyoto-u.ac.jp, nojiri@nda.ac.jp}}
\address{Department of Applied Physics,
National Defence Academy,
Hashirimizu Yokosuka 239-8686, JAPAN}

\author{Sergei D. Odintsov\footnote{E-mail: 
odintsov@ieec.fcr.es Also at TSPU, Tomsk, Russia}}
\address{Instituci\`o Catalana de Recerca i Estudis Avan\c{c}ats (ICREA)
and Institut d'Estudis Espacials de Catalunya (IEEC), 
Edifici Nexus, Gran Capit\`a 2-4, 08034 Barcelona, SPAIN}

\begin{abstract}
It is proposed that dark energy may become dominant over standard 
matter due to universe expansion (curvature decrease). Two  
models: non-linear gravity-matter system and modified gravity may 
provide an effective phantom or effective quintessence dark energy  
which complies with the conjecture. 
The effective quintessence naturally  describes current cosmic speed-up.
\end{abstract}


\maketitle

\noindent
{\bf 1. Introduction.} 
Growing evidence from high redshift surveys of supernovae, WMAP data
analysis and other sources indicates that current universe expiriences a 
cosmic speed-up phase. Moreover, about 70 percent of total universe energy 
is attributed to puzzling cosmic fluid with large negative pressure (dark
energy). The simplest possibility for dark energy is a cosmological
constant (for recent review, see \cite{alexei,peebles}) with equation of
state
parameter $w=-1$. Unfortunately, the dark energy due to cosmological
constant requires huge fine-tuning. Moreover, it could be that the
realistic value for $w$ is slightly less than $-1$. Such phantom dark
energy may be described by scalar field with negative kinetic energy
\cite{phantom}. Unfortunately, phantoms look inconsistent in many
respects. In
particulary, phantom energy becomes infinite in finite time and such
universe quickly evolves to Big Rip \cite{brett}. (Note, however, that
finite
time future singularity is possible even if strong-energy condition holds
\cite{john}). The number of other proposals for dark energy exists
(for a review, see \cite{p}) but most
of them have also serious drawbacks. As a rule, the dark energy models 
do not respond to the question: why dark energy became the dominant
contribution to energy density  
precisely at current epoch?

Recently, a gravitational alternative for dark energy was suggested
\cite{turner} modifying the Einstein action by adding a $1/R^n$  term.
Such a theory may produce the current cosmic speed-up, may be naturally
generated by M-theory \cite{NO1} and may give an effective phantom
description with $w$ above (or less) than $-1$. Of course, higher 
derivative gravity contains a number of instabilities \cite{ins}
(for a recent discussion of possible deviations from Newton gravity, see \cite{lue}). 
Nevertheless, some modification of such a theory at high curvature (by
higher derivatives and $\ln$-terms) \cite{NO2} leads to viable 
theory (with some amount of fine-tuning). This picture is supported by
quantum effects account \cite{NO2}. It is interesting that Palatini version 
of modified gravity may be also consistent \cite{palatini}. Of course, 
as deviations from GR occur at low curvatures (for instance, in entropy
studies \cite{brevik}), more checks of such modified gravity should be done.
Nevertheless, the important lesson drawn by such modification is 
possible, very simple explanation of current dark energy dominance.
Indeed, gravitational dark energy is  increased due to dynamical decrease 
of the curvature while FRW universe expands. It is extremely interesting
to understand if such natural explanation may be applied to another dark
energy models where probably gravity itself is less modified. 

In the present letter we make some attempt in the construction of dark
energy which grows due to decrease of the curvature. Such gravity assisted
 dark energy dominance may occur due to 
direct (non-linear) gravitational coupling with matter-like lagrangian
as we show below.   

\noindent
{\bf 2. Non-linear matter-gravity coupling as asymptotic dark energy and
cosmic speed-up.} 
Let us start from the following action:
\be
\label{LR1}
S=\int d^4 x \sqrt{-g}\left\{{1 \over \kappa^2}R + \left(\frac{R}{\mu^2}\right)^\alpha L_d \right\}\ .
\ee
Here $L_d$ is matter-like action (dark energy). The choice of  parameter 
$\mu$ may keep away the unwanted instabilities which often occur in higher 
derivative theories.
The second term in above action describes the non-linear coupling of
matter with gravity (in parallel with  $R \phi^2$ term which is 
usually required by renormalizability condition).
 Similarly, such term may be induced by quantum
effects as non-local effective action. Hence, it
is natural to consider that it belongs to matter sector. (Standard matter 
is not included for simplicity).
It is also interesting to remark that higher derivative kinetic term of 
above sort was proposed  in \cite{MR} for the study of cosmological 
constant problem. However, this model is different from our theory
because it also 
contained $R^2$ term in gravitational sector while the non-linear coupling 
of the potential with the curvature was not included.

By the variation over $g_{\mu\nu}$, the equation of motion follows:
\be
\label{LR2}
0= {1 \over \sqrt{-g}}{\delta S \over \delta g_{\mu\nu}} 
= {1 \over \kappa^2}\left\{{1 \over 2}g^{\mu\nu}R - R^{\mu\nu}\right\} 
+ \tilde T^{\mu\nu}\ .
\ee
Here  the effective 
energy momentum tensor (EMT) $\tilde T_{\mu\nu}$ is defined by
\bea
\label{w5}
\tilde T^{\mu\nu}&\equiv& \frac{1}{\mu^{2\alpha}}\left\{ - \alpha R^{\alpha - 1} R^{\mu\nu} L_d \right. \nn
&& \left. + \alpha\left(\nabla^\mu \nabla^\nu 
 - g^{\mu\nu}\nabla^2 \right)\left(R^{\alpha -1 } L_d\right) + R^\alpha T^{\mu\nu}\right\}\ ,\nn
T^{\mu\nu}&\equiv& {1 \over \sqrt{-g}}{\delta \over \delta g_{\mu\nu}}
\left(\int d^4x\sqrt{-g} L_d\right) 
\eea
In accord with our last remark, we consider last term in equation
of motion as the effective matter EMT. For scalar matter
below one may insist that above theory is just scalar-tensor gravity of
unusual form (modified gravity). Nevertheless, other matter (fermion, Yang-Mills field)
may be considered too. Then, the interpretation of above theory as modified
gravity is doubtful.

Let free massless scalar be a matter
\be
\label{LR4}
L_d = - {1 \over 2}g^{\mu\nu}\partial_\mu \phi \partial_\nu \phi\ .
\ee
The metric is chosen to describe the FRW universe with flat 3-space:
\be
\label{LR6}
ds^2 = - dt^2 + a(t)^2 \sum_{i=1,2,3} \left(dx^i\right)^2\ .
\ee

If we assume $\phi$ only depends on $t$ $\left(\phi=\phi(t)\right)$, the solution of 
scalar field equation is given by 
\be
\label{LR9}
\dot \phi = q a^{-3} R^{-\alpha}\ .
\ee
Here $q$ is a constant of the integration. Hence $R^\alpha L_d = {q^2 \over 2 a^6 R^\alpha}$, 
which becomes dominant when $R$ is small (large) compared with the 
Einstein term ${1 \over \kappa^2}R$ if $\alpha>-1$ $\left(\alpha <-1\right)$.
Thus, one arrives at the remarkable possibility that dark energy grows
to asymptotic dominance over the usual matter with decrease of the
curvature.


Substituting (\ref{LR9}) into (\ref{LR2}),  the $(\mu,\nu)=(t,t)$
component of equation of motion  
has the following form:
\bea
\label{LR11}
&& 0=-{3 \over \kappa^2}H^2 + \frac{36q^2}{\mu^{2\alpha} a^6\left(6\dot H + 12 H^2\right)^{\alpha + 2}} \nn
&& \times\left\{ {\alpha (\alpha + 1) \over 4}\ddot H H + {\alpha + 1 \over 4}{\dot H}^2 
\right. \nn
&& \left. + \left(1 + {13 \over 4}\alpha + \alpha^2\right)\dot H H^2 
+ \left(1 + {7 \over 2}\alpha\right) H^4 \right\}\ .
\eea
Especially when $\alpha = -1$, 
this equation has only the trivial solution $H=0$ ($a$ is constant). 

The accelerating solution of (\ref{LR11}) exists
\bea
\label{LR13}
&& a=a_0 t^{\alpha + 1 \over 3}\quad \left(H={\alpha + 1 \over 3t}\right)\ ,\nn
&& a_0^6 \equiv \frac{\kappa^2 q^2 \left(2\alpha - 1\right)\left(\alpha - 1\right) }
{\mu^{2\alpha}  3\left(\alpha + 1\right)^{\alpha + 1}
\left({2 \over 3}\left(2\alpha - 1\right)\right)^{\alpha + 2}}\ .
\eea
Eq.(\ref{LR13}) tells that the universe accelerates, that is, $\ddot a>0$
if $\alpha>2$. If $\alpha<-1$, the solution (\ref{LR13}) describes shrinking universe. But if we change the 
direction of the time as $t\to t_s - t$ ($t_s$ is a constant), we have accelerationg and expanding 
universe. In the solution with $\alpha<-1$ there appears a singularity at $t=t_s$, which is 
Big Rip singularity.   
For the matter with the relation $p=w\rho$, where $p$ is the pressure and $\rho$ is the 
energy density, from the usual FRW equation, one has $a\propto t^{2 \over 3(w+1)}$.
For $a\propto t^{h_0}$ it follows $w=-1 + {2 \over 3h_0}$, 
and the accelerating expansion ($h_0>1$) of the universe occurs if
$-1<w<-{1 \over 3}$.
For the case of (\ref{LR13}), we find $w={1 - \alpha \over 1 + \alpha}$. 
Then if $\alpha<-1$, we have $w<-1$, which is an effective phantom.
For the general matter with the relation $p=w\rho$ with constant $w$, 
the energy $E$ and the energy density $\rho$ behave as
$E\sim a^{-3w}$ and $\rho\sim a^{-3\left(w + 1\right)}$.
Thus, for the standard phantom with $w<-1$, the density becomes large with 
time and might generate finite time future singularity  (Big Rip). 



For the solution (\ref{LR13}), the first and second terms in (\ref{LR1}) are of the same order. 
This is true for both of the early time (small $t$) and late time (large $t$) epochs. 
Let us take the case that the second term  is dominant and  the first term may be neglected. 
Assuming $H={h_0 \over t}$, we find
$h_0 = {1 + {13 \over 4}\alpha + \alpha^2 \pm \sqrt{\alpha^4 - {\alpha^3 \over 2}} 
\over 2\left(1 + {7 \over 2}\alpha\right)}$, 
which is real if $\alpha\leq 0$ or $\alpha>{1 \over 2}$. Since 
$R^\alpha L_d = {q^2 \over 2 a^6 R^\alpha}$, the second 
term in (\ref{LR1}) behaves as $R^\alpha L_d \sim t^{2\alpha - 6h_0} = 
t^{-3 - {31 \over 4}\alpha + 4\alpha^2 \mp \sqrt{\alpha^4 - {\alpha \over 2}} 
\over 1 + {7 \over 2}\alpha}$. Especially if $\alpha$ is large,  
$R^\alpha L_d \sim t^{2\alpha - 6h_0} = t^{{2 \over 7}\alpha,\ 2\alpha}$.
As the scalar curvature behaves as $R\sim t^{-2}$, if $\alpha$ is positive and large, 
the second term in (\ref{LR1}) becomes surely dominant compared with the first 
Einstein term when $R$ is small, that is, $t$ is large. 
In the early  universe (small $t$), the second term might be suppressed. 
Thus, if   $w$ is  negative but bigger than
$-1$ in the current universe, the model under discussion (effective
quintessence\cite{quint}) describes the 
current cosmic speed-up. On the same time, such dark energy dominance is 
again explained by the universe expansion. One can also show that 
corrections to Newton law in this scenario are small.  


In \cite{ENOprd}, the stability of the solution (\ref{LR13}) has been investigated by replacing 
$a=at^{\frac{\alpha+1}{3}}$ with $a=at^{\frac{\alpha+1}{3}}\left(1+\delta\right)$ $\left(|\delta|\ll 1\right)$. 
There $\delta$ has been assumed to only depend on the time variable $t$. 
Then the equation with account of perturbations is
\be
\label{P1}
0=\frac{1}{t}\frac{d^3\delta}{dt^3} + \frac{A_1}{t^2}\frac{d^2\delta}{dt^2} 
+ \frac{A_2}{t^3}\frac{d\delta}{dt} + \frac{A_3}{t^4}\delta\ ,
\ee
with constants $A_1$, $A_2$, $A_3$, and $A_4$. 
One may consider the case that $\delta$ depends on the spatial coordinates.
Since this dependence 
appears through the d'Alembertian, after replacing the Laplacian with $-k^2$ ($k$ is the magnitude of the 
momentum), Eq.(\ref{P1}) should be modified as
\be
\label{P2}
0=\frac{1}{t}\frac{d^3\delta}{dt^3} + \frac{A_1}{t^2}\frac{d^2\delta}{dt^2} 
+ \frac{A_2}{t^3}\frac{d\delta}{dt} + \frac{A_3}{t^4}\delta
+ B_1\frac{k^2}{t^2} \frac{d^2\delta}{dt^2} + \frac{B_2k^2 }{t}\frac{d\delta}{dt} \ ,
\ee
with constants $B_1$ and $B_2$. The newly added terms in (\ref{P2}) may be dominant when $t$ is large 
but less dominant when $t$ is small. When $w<-1$, after replacing $t$ by $t_0 -t$, small $t$ corresponds 
to the case $t\sim t_0$. Then the inhomogeneity of the universe does not grow up when $w<-1$. 



One can rewrite the action (\ref{LR1}) by using the auxilliary field(s). 
First we introduce two scalar field $\zeta$ and $\eta$ and rewrite (\ref{LR1})
as
\be
\label{LR18}
S=\int d^4 x \sqrt{-g}\left\{{1 \over \kappa^2}\zeta + \zeta^\alpha L_d 
+ \eta \left(R - \zeta\right) \right\}\ .
\ee
Using the equation $\zeta=R$ given by the variation over $\eta$,
the action  (\ref{LR18}) is reduced into the original one  (\ref{LR1}).
Varying over 
$\zeta$, we obtain
$\eta = {1 \over \kappa^2} + \alpha \zeta^{\alpha - 1}L_d$. 
For $\alpha\neq 1$, one can delete $\zeta$ in (\ref{LR18}) as
\bea
\label{LR21}
S&=&\int d^4 x \sqrt{-g}\left\{\eta R + \left({1 \over \alpha}-1\right) \right.\nn
&& \left. \times \left(\eta - 
{1 \over \kappa^2}\right)^{1 \over 1 - \alpha}
\left(\alpha L_d \right)^{1 \over 1-\alpha}\right\}\ .
\eea
Writing $\eta$ as $\eta=\e^{-\sigma}$ and rescaling the metric as $g_{\mu\nu}\to \e^\sigma g_{\mu\nu}$, 
 the Einstein frame action follows:
\bea
\label{LR21b}
S&=&\int d^4 x \sqrt{-g}\left\{R - \frac{3}{2}g^{\mu\nu}\partial_\mu\sigma \partial_\nu\sigma 
+ \left({1 \over \alpha}-1\right) \right.\nn
&& \left. \times
\left(\e^{-\sigma} - 
{1 \over \kappa^2}\right)^{1 \over 1 - \alpha}
\left(\alpha L_d\left(\e^\sigma g_{\mu\nu}, \phi\right) \right)^{1 \over 1-\alpha}\right\}\ .
\eea
Such the non-linear action includes Brans-Dicke type scalar $\sigma$ and the scalar $\phi$ 
corresponding to the dark energy, that is,  two scalars appear.
However, as is expected the (non-standard) kinetic term for $\phi$ describes the coupling with $\sigma$ on the same time.
Some remark about the equivalence principle may be in order. 
First, note that there is a trivial solution with $q=0$ in (\ref{LR6}), where $H=R=0$ 
and $\phi=0$. Since the curvature in the present universe is small, one may assume 
$H$, $R$, $\phi\sim 0$. What about the perturbation around the solution $R=H=\phi=0$ 
in the action (\ref{LR1})? If $\alpha>0$, there does not appear $\phi$ in
the perturbed action. Hence, if 
 the usual matter action  does not couple with $\phi$ directly, the equivalence principle is not 
violated.

Immediate generalization of (\ref{LR4}) is to include the potential:
\be
\label{LR22}
L_d = - {1 \over 2}g^{\mu\nu}\partial_\mu \phi \partial_\nu \phi - V(\phi)\ .
\ee
The solution may be found for a special choice (as an illustrative example) of
$V(\phi) = V_0 \phi^{2 - {2 \over \alpha}}$ with a constant $V_0$,  
if we assume the FRW metric (\ref{LR6}) and if $\alpha\neq 1$ and 
$\alpha \neq - 1 + 3h_0$:
\be
\label{LR25}
\phi = \phi_0 t^\alpha\ ,\quad H={h_0 \over t}\ \quad \left(a=a_0 t^{h_0}\right)\ .
\ee
Here the constants $\phi_0$ and $h_0$ are expressed in terms of the theory
parameters.
 If $\alpha=-1 + 3h_0<0$, there is only trivial 
solution with $\phi_0 = h_0 =0$. On the other hand, if $\alpha=-1 + 3h_0>0$ and $V_0\neq 0$, 
there is no solution. 

For the solution (\ref{LR25}) with dominant second term in (\ref{LR1}) 
 and putting ${1 \over \kappa^2}\to 0$
one gets $h_0 = {\alpha -3 \over 3\left(\alpha - 2\right)}$. 
Hence, if ${3 \over 2}<\alpha < 2$,  $h_0>0$, what may correspond
to the cosmic speed-up with $w={\alpha - 1 \over \alpha - 3}$. 
This is an effective quintessence dark energy.
If $1<\alpha<2$, we obtain an effective phantom with $w<-1$.
In both cases, the current dark energy dominance is explained by 
the universe expansion. 
Similarly, one can analyze the potentials of other form. 

As the generalization of other type we consider the model:
\be
\label{gLR1}
S=\int d^4 x \sqrt{-g}\left\{{1 \over \kappa^2}R + f(R) L_d \right\}\ .
\ee
Assuming that $f(R)$ behaves as $f(R)\sim R^\alpha$ when $R$ is small and 
$f(R)\sim R^\beta$ when $R$ is large, as an example,  one can take 
$f(R) = a  R^\alpha + b R^\beta$, 
with $\alpha<\beta$.  $L_d$ is chosen in the form (\ref{LR4}). When
$R$ is small, we obtain (\ref{LR13}) again.
On the other hand, when $R$ is large:
\be
\label{gLR4}
a\propto t^{\beta + 1 \over 3}\ ,\quad w\sim {1 - \beta \over 1 - \beta}\ .
\ee
If $\beta> -1$, at the early time, the universe expands as (\ref{gLR4}). 
With the growth of time, $R\sim t^{-2}$ and when it
 becomes sufficiently small, the universe accelerates. 
Thus, the possibility to unify the early time inflation with current 
(asymptotic) dark energy dominance appears.

\noindent
{\bf 3. Modified gravity with time-dependent coefficients.} 
Another class of models may be considered:
\be
\label{LR28}
S=\int d^4 x \sqrt{-g}\left({R \over \kappa^2} + {\alpha(t) \over R} + \beta(t)\right)\ .
\ee
Here $\alpha(t)$ and $\beta(t)$ are time-dependent
phenomenological parameters. The above non-covariant action may be 
considered as an effective theory in parallel with the effective (non-covariant) theory 
of time-dependent cosmological constant. From another side, it may origin from 
more fundamental covariant gravity-matter system of the sort described 
in previous section\footnote{
The action (\ref{LR28}) may origin from more complicated, non-linear action like 
$S=\int d^4 x \sqrt{-g}\left({R \over \kappa^2} + {L_\alpha \over R} + L_\beta\right)$. 
In fact one may choose $L_\alpha$ as in (\ref{LR22}), $L_\alpha = 
- {1 \over 2}g^{\mu\nu}\partial_\mu \phi \partial_\nu \phi - V_{0\alpha} \phi^4$, which corresponds to $\alpha=-1$ and 
$L_\beta$ as $L_\beta = - {1 \over 2}\partial_\mu \varphi \partial_\nu \varphi 
 - V_{0\beta} \e^{-2\frac{\varphi}{\varphi_0}}$. Then after solving 
 FRW equations, we find 
$L_\alpha\propto t^{-4}$ and $L_\beta\propto t^{-2}$. It is interesting that such model may 
be rewritten in the Einstein frame (similar to Eq.(20)) as two scalars-tensor gravity. 
}.
In the FRW metric (\ref{LR6}),
the simple solution of equation of motion is $H={h_0 \over t}$,
$\alpha={\alpha_0 \over t^4}$,
and 
$\beta={\beta_0 \over t^2}$. 
Here the constants  $h_0$, $\alpha_0$, and $\beta_0$ are related by
\be
\label{LR32}
0=-{3h_0^2 \over \kappa^2} - {\beta_0 \over 2} - {\left(3h_0 - 2\right)\alpha_0 
\over 12\left(2h_0 - 1\right)^2 h_0}\ .
\ee
Eq.(\ref{LR32}) tells that $\alpha_0$ and/or $\beta_0$ should be 
negative when $h_0>{2 \over 3}$.
Since $a \propto t^{h_0}$, the cosmic acceleration $\left(\dot a>0,\ \ddot a>0\right)$ 
occurs if $h_0>1$.  
 $w\sim -1$ corresponds to the limit of 
$\left|h_0\right|\to \infty$. In case $\left|h_0\right|\gg 1$,
in order that $h_0$ is real, one gets $\alpha_0<0$. 
Since $w=-1 + {2 \over 3h_0}$, $w$ can be found as
\be
\label{LR37}
w=-1 \pm {2 \over 3\sqrt{ - \alpha_0 \kappa^2}}\sqrt{1 + \sqrt{1 - {3\alpha_0 \over 
\beta_0^2 \kappa^2}}}\ .
\ee
Then the plus sign in (\ref{LR37}) corresponds to quintessence and the minus one 
to phantom. Again the interpretation of dark energy dominance due to the
curvature decrease is immediate.

The action (\ref{LR28}) may be rewritten in the form of the scalar-tensor
theory. 
First by using the auxiliary field $\sigma$, the action (\ref{LR28}) is
\be
\label{LR38}
S=\int d^4 x \sqrt{-g}\left\{\left(1 - {\kappa^2\sigma^2 \over 4 \alpha(t)}\right)
{R \over \kappa^2} + \sigma + \beta(t)\right\}\ .
\ee
By defining new field $\varphi$ as 
$\e^{-\varphi}\equiv 1 - {\kappa^2\sigma^2 \over 4 \alpha(t)}$, 
we rescale the metric tensor as $g_{\mu\nu}\to \e^\varphi g_{\mu\nu}$. 
Then the action (\ref{LR38}) describes scalar-tensor theory with
time-dependent potential:
\bea
\label{LR42}
S&=&\int d^4 x \sqrt{-g}\left\{ {1 \over \kappa^2 }\left( R - {3 \over 2}\partial_\mu \varphi 
\partial^\mu \varphi \right)\right. \nn
&& \left. \pm {2 \over \kappa} \e^{2\varphi}
\sqrt{\alpha(t)\left( 1 - \e^{-\varphi}
\right)} + \e^{2\varphi}\beta(t)\right\}\ .
\eea
This indicates that such models of dark energy dominance are in different
class (more close to modified gravity \cite{turner,NO2}) with non-linear 
gravity-matter theory of previous section. They may contain the
instabilities \cite{ins}.

In summary, gravity assisted dark energy (the possible origin of cosmic
speed-up) may become dominant over standard matter just because of 
the universe expansion (curvature decrease). 
It is a challenge to understand if such conjecture is true 
and if one of simple models proposed in this work may lead to
 dark energy theory which complies with observational data.

\noindent
{\bf Acknowledgments} 
This investigation was supported in part by the Monbusho of Japan
under the grant n.13135208
(S.N.).


\begin{thebibliography}{99}

\bibitem{alexei} V. Sahni and A. Starobinsky, {\sl Int.J.Mod.Phys.}
{\bf D9} (2000) 373, astro-ph/9904398.
\bibitem{peebles} P.J. Peebles and B. Ratra, {\sl Rev.Mod.Phys.}
{\bf 75} (2003) 599, astro-ph/0207347.
\bibitem{phantom} R.R. Caldwell, {\sl Phys.Lett.} {\bf B545} (2002)  23-29, astro-ph/9908168; 
A.E. Schulz and M.J. White, {\sl Phys.Rev.} {\bf D64} (2001) 043514, astro-ph/0104112; 
T. Chiba, T. Okabe and M. Yamaguchi, {\sl Phys.Rev.} {\bf D62} (2000) 023511;  
V. Faraoni, {\sl IJMP} {\bf D64} (2002) 043514; 
G.W. Gibbons, hep-th/0302199; 
S. Nojiri and S.D. Odintsov, {\sl Phys.Lett.} {\bf B562} (2003) 147, hep-th/0303117;  
{\sl Phys.Lett.} {\bf B571} (2003) 1, hep-th/0306212; 
 {\sl Phys.Lett.} {\bf B565} (2003) 1, hep-th/0304131; 
P. Singh, M. Sami and N. Dadhich, {\sl Phys.Rev.} {\bf D68} (2003) 023522, hep-th/0305110; 
J.-g. Hao and X.-z. Li, hep-th/0303093; astro-ph/0309746; 
M.P. Dabrowski, T. Stachowiak and M. Szydlowski, hep-th/0307128; 
L.P. Chimento and R. Lazkoz, gr-qc/0307111, {\sl Phys.Rev.Lett.} {\bf 91} (2003)211301; 
A. Feinstein and S. Jhingan, hep-th/0304069; 
P.F. Gonzalez-Diaz, astro-ph/0305559; astro-ph/0312579; 
E. Elizalde and J.Q. Hurtado, gr-qc/0310128; 
Y. Piao and E. Zhou, hep-th/0308080; 
H. Stefancic, astro-ph/0310904; astro-ph/0312484; 
M. Sami and A.Toporensky, gr-qc/0312009; 
J. Cline, S. Jeon and G. Moore,hep-ph/0311312; 
X. Meng and P. Wang,hep-ph/0309746; H.Q. Lu, hep-th/0312082; 
V.B. Johri, astro-ph/0311293;
J.S. Alcaniz, astro-ph/0312424;
G. Calcagni, hep-ph/0402126;
J. Aguirregabiria, L.P. Chimento and R. Lazkoz, astro-ph/0403157. 
\bibitem{brett} B. McInnes, {\sl JHEP} {\bf 0208} (2002) 029;
A.A. Starobinsky, {\sl G@C} {\bf 6} (2000) 157;
R.R. Caldwell, M. Kamionkowski and N.N. Weinberg,{\sl Phys.Rev.Lett.} {\bf 91} (2003) 071301, astro-ph/0302506.
\bibitem{john} J.D. Barrow, gr-qc/0403084. 
\bibitem{p} T. Padmanabhan, {\sl Phys.Repts.} {\bf 380} (2003)235.
\bibitem{turner} S.M. Carroll, V. Duvvuri, M. Trodden and M.S. Turner, astro-ph/0306438;
S. Capozzielo, S. Carloni and A. Troisi, astro-ph/0303041.
\bibitem{NO1} S. Nojiri and S.D. Odintsov,{\sl Phys.Lett.} {\bf B576} (2003)5, hep-th/0307071.
\bibitem{ins} T. Chiba, astro-ph/0307338;
A.D. Dolgov and M. Kawasaki, astro-ph/0307285;
M.E. Soussa and R.P. Woodard, astro-ph/0308114;
A. Nunez and S. Salganik, hep-th/0403159.
\bibitem{lue}A. Lue, R. Scoccimarro and G.D. Starkman, astro-ph/0401515.
\bibitem{NO2} S. Nojiri and S.D. Odintsov,{\sl Phys.Rev.} {\bf D68} (2003)123512, hep-th/0307288;
{\sl GRG} {\bf 36} (2004) 1765, hep-th/0308176; {\sl Mod.Phys.Lett.} {\bf A19} (2004) 627, hep-th/0310045.
\bibitem{palatini} D. Vollick,{\sl Phys.Rev.} {\bf D68} (2003) 063510, astro-ph/0306630; 
X. Meng and P. Wang, astro-ph/0307354; astro-ph/0308031; astro-ph/0308284;
E. Flanagan, {\sl Phys.Rev.Lett.} {\bf 92} (2004)071101, astro-ph/0308111; gr-qc/0403063;
Y. Ezawa et al, gr-qc/0309010;
G. Allemandi, A. Borowiec and M. Francaviglia, hep-th/0403264;
G. Olmo and W. Komp, gr-qc/0403092.
\bibitem{brevik}  I. Brevik, S. Nojiri, S.D. Odintsov and L. Vanzo, to appear {\sl Phys.Rev.} {\bf D}, hep-th/0401073. 
\bibitem{MR}  S. Mukohyama and L. Randall, {\sl Phys.Rev.Lett.} {\bf 92} (2004) 211302, hep-th/0306108; 
A.D. Dolgov and M. Kawasaki, astro-ph/0307442, astro-ph/0310822.  
\bibitem{quint} B. Ratra and P.J.E. Peebles, {\sl Phys.Rev.} {\bf D37} (1988) 3406;
R.R. Caldwell, R. Dave and P.J. Steinhardt, {\sl Phys.Rev.Lett.} {\bf 80}  (1998) 1582;
N. Bahcall, J.P. Ostriker, S. Perlmutter and P.J. Steinhardt, {\sl Science} {\bf 284} (1999) 1481.
\bibitem{ENOprd} E. Elizalde, S. Nojiri, S.D. Odintsov, tp appear in {\sl Phys.Rev.} {\bf D}, hep-th/0405034. 

\end{thebibliography}
\end{document}